\documentclass[manuscript,screen]{acmart}
\AtBeginDocument{%
  \providecommand\BibTeX{{%
    \normalfont B\kern-0.5em{\scshape i\kern-0.25em b}\kern-0.8em\TeX}}}
    
\usepackage[utf8]{inputenc} 
\usepackage[T1]{fontenc}    
\usepackage{hyperref}       
\usepackage{url}            
\usepackage{booktabs}       
\usepackage{amsfonts}       
\usepackage{nicefrac}       
\usepackage{microtype}      
\usepackage{lipsum}
\usepackage[linesnumbered,algoruled,boxed,lined]{algorithm2e}
\usepackage{amsmath}
\usepackage{subfigure}

\usepackage{graphicx}       
\graphicspath{{figures/}}

\copyrightyear{2019} 
\acmYear{2019} 
\setcopyright{rightsretained} 
\acmConference[SpatialGems '19]{1st ACM SIGSPATIAL International Workshop on Spatial Gems}{November 5--8, 2019}{Chicago, Illinois, USA}
\acmBooktitle{1st ACM SIGSPATIAL International Workshop on Spatial Gems (SpatialGems '19), November 5--8, 2019, Chicago, Illinois, USA}
\acmDOI{10.1145/3391234.3421234}
\acmISBN{978-1-4503-8019-5/20/11}

\begin{document}

\title{Spatial Data Generators}

\author{Tin Vu}
\affiliation{
\institution{Dept. of Computer Science and Engineering,
University of California, Riverside}
\city{Riverside}
\state{CA}
\country{USA}
}
\email{tin.vu@email.ucr.edu}

\author{Alberto Belussi}
\affiliation{
\institution{Dept. of Computer Science,
University of Verona}
\city{Verona}
\country{Italy}
}
\email{alberto.belussi@univr.it}

\author{Sara Migliorini}
\affiliation{
\institution{Dept. of Computer Science,
University of Verona}
\city{Verona}
\country{Italy}
}
\email{sara.migliorini@univr.it}

\author{Ahmed Eldawy}
\affiliation{
\institution{Dept. of Computer Science and Engineering,
University of California, Riverside}
\city{Riverside}
\state{CA}
\country{USA}
}
\email{eldawy@ucr.edu}

\renewcommand{\shortauthors}{Tin Vu, et al.}

\begin{abstract}
This gem describes a standard method for generating synthetic spatial data that can be used in benchmarking and scalability tests. The goal is to improve the reproducibility and increase the trust in experiments on synthetic data by using standard widely acceptable dataset distributions. In addition, this article describes how to assign a unique identifier to each synthetic dataset that can be shared in papers for reproducibility of results. Finally, this gem provides a supplementary material that gives a reference implementation for all the provided distributions.
\end{abstract}




\maketitle

\section{Introduction}

In many published papers, researchers often need to test their implementations of new index structures or query execution methods on large scale spatial data. While some real datasets exist, the research community also needs to try datasets with specific characteristics to highlight how the proposed research behaves under certain circumstances. Synthetic data generation gives researchers full control over the data characteristics such as data skewness, complexity of geometries, or amount of overlap between datasets.

This article proposes a practical tool for generating synthetic spatial datasets with various skewed distributions. These generators have been successfully used in existing research to evaluate index construction, query processing, spatial partitioning, and cost model verification. While the generators are already used in many papers, there is a fact that the researchers rarely describe the details of generating these datasets for two reasons. First, it is not usually a research contribution and the authors do not want to draw attention to it. Second, it takes a precious space of the paper that authors usually prefer to utilize for other parts.

This gem takes the burden of describing, in detail, how to generate synthetic data of six common distributions. As the gems are designed to be flexible, it fits very well the generation of synthetic data where other researchers can add more datasets in the future.

\begin{figure}
\centering
\includegraphics[width=\textwidth]{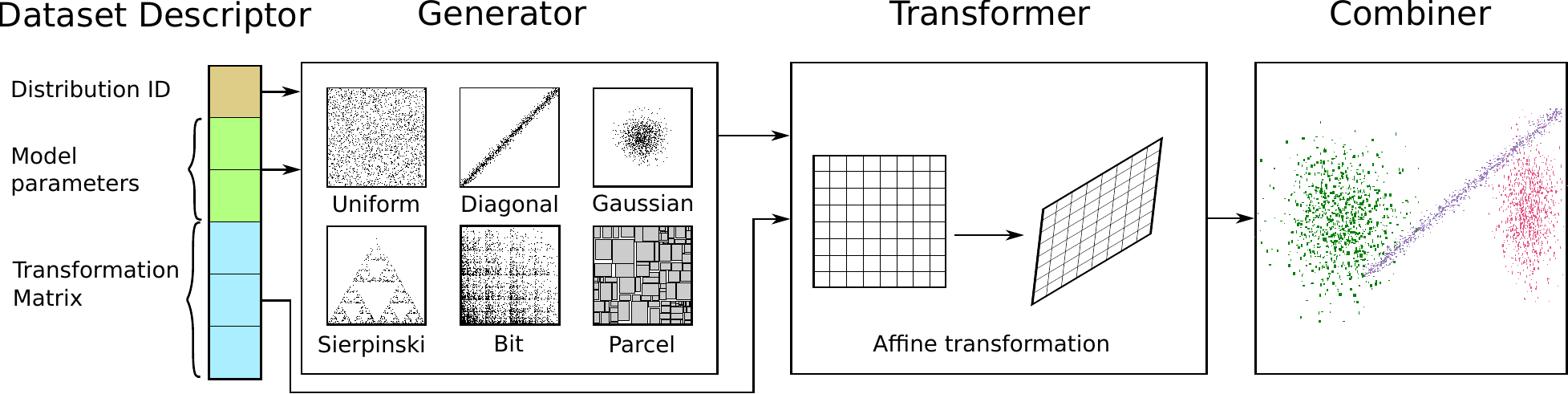}
\caption{Overview of the spatial generator}
\label{fig:overview}
\end{figure}

Figure~\ref{fig:overview} gives an overview of the main parts of the proposed generator. First, the dataset descriptor is a vector that contains information about the dataset to be generated. It acts a unique identifier for the synthetic dataset and it consists of three parts,
(1)~the distribution ID $i\in[1,6]$ for six implemented distributions,
(2)~the model parameters depending on the chosen distribution, and
(3)~a transformation matrix used later by the transformer.
The first two components of the dataset descriptor, i.e., distribution ID and model parameters, are passed to the generator which generates the desired dataset. After that, the transformer applies an affine transformation on the generated data according to the third component of the dataset descriptor.

Since the dataset descriptor fully identifies the generated dataset, researchers who use these generators can simply cite this gem and list the descriptors of the datasets they used. Other researchers can then regenerate the same datasets following the same procedure described in this gem. For guidance, it provides a reference implementation~\cite{github-sdg} to all the generators as a supplementary material, but researchers can develop other generators that follow the same guidelines, e.g., a Spark-based generator for big spatial data. For example, Table~\ref{tab:datasets} at the end of this gem defines the six datasets used in this article.

The final component, combiner, can be used to create compound datasets by simply merging two or more simple datasets. In this case, the descriptor of the compound dataset is simply the concatenation of all the descriptors of the simple datasets.

As shown in Figure~\ref{fig:overview}, this gem uses six different distributions for generating the simple datasets, which are all described in the next section.

\section{Data Generators}
This section describes the six synthetic generators which are used in this gem, namely, uniform, diagonal, Gaussian, Sierpinski, bit, and parcel. Some of these generators are inspired by a benchmark developed by Beckmann and Seeger~\cite{BS}. This article provides more details about these generators and defines some additional generators.
Each generator $G_*$ takes a list of common parameters $[cp_1, cp_2, \dots]$ and a list of distribution-specific parameters $[sp_1, sp_2, \dots]$. For the family of generators that are considered in this gem, there are two common parameters, the \textsc{dataset cardinality} ($card$) specifying the total number of geometries and the \textsc{number of dimensions} ($d$). The generator here considers the generation of two-dimensional geometries; the extension to multi-dimensional datasets is straightforward.

For all these generators, the \textsc{reference space} that contains the generated data is $[0,1]^d$, where $d$ is the number of dimensions. Additionally, these generators assume the existence of a random number generator {\sc Rnd()} which generates random numbers in the range $[0,1)$. This generator can be used to generate random numbers for three popular distributions, Bernoulli, Uniform, and Normal, as follows~\footnote{More efficient implementations are usually available in standard packages}.

\begin{equation}
Bernoulli(p)=\left\{
\begin{array}{cc}
     1& ;\textsc{Rnd()} < p \\
     0& ; otherwise
\end{array}
\right.
\end{equation}

\begin{equation}
    U(a,b)=(b-a)\textsc{Rnd()} + a
\end{equation}

\begin{equation}
    N(\mu,\sigma)=\mu+\sigma\sqrt{-2\ln \textsc{Rnd}_1()} \cdot \sin(2\pi \textsc{Rnd}_2()) (\text{Box and Muller~\cite{box1958}})
\end{equation}

Algorithm~\ref{algo:basicGen} shows the general schemata of the first five generators, namely, uniform, diagonal, Gaussian, Sierpinski, and bit distribution. These five generators can generate both points and rectangles. Points are generated using the GenPNT$_*$ methods, described shortly, while rectangles are generated by using these points as centers. Lines~\ref{algo:basicGen:init1},\ref{algo:basicGen:init2} initialize the result set $\mathcal G$ to the empty set and the random number generator. If desired, the seed of the random number generator can be fixed to generate exactly the same random dataset. The loop in lines~\ref{algo:basicGen:loop1}-\ref{algo:basicGen:loop2} generates one record at a time. Line~\ref{algo:basicGen:genpnt} calls a generic {\sc GenPNT$_*$} function that is different for each generator. This function returns a random point $(x,y)$ according to the desired distribution. Then, Line~\ref{algo:basicGen:check} tests if the point is inside the reference space $[0,1]^d$ as some generators can generate points outside that space, e.g., Gaussian. If the point is inside the reference space, the algorithm continues by generating random width and height for the rectangle by using the parameters $sp_1$ and $sp_2$ as the maximum allowed width and height. Finally, a rectangle is generated with $(x,y)$ as the center and $(w,h)$ as its dimensions, i.e., the corner point is at $(x-w/2,y-h/2)$. The notation $Box(x,y,w,h)$ is used to indicate a box with its lower corner point at $(x,y)$ and has dimensions of $w$ and $h$.

The following parts are the descriptions of the point generators ({\sc GenPNT$_*$}) for the first five distributions and then the parcel distribution which generates rectangles directly without generating points first.

\begin{algorithm}[htb]
\SetAlgoLined
\KwIn{$card, d=2, sp_1, \dots, sp_n$}
\KwResult{Set of geometries: $\mathcal{G}=\{ geom \}$}
$\mathcal{G} \leftarrow \emptyset$; $i \leftarrow 0$\; \label{algo:basicGen:init1}
Initialize the random number generator\; \label{algo:basicGen:init2}
\While{$i < card$} {\label{algo:basicGen:loop1}
       $(x,y) \leftarrow \textsc{GenPNT}_*(i, sp_3, \dots, sp_n)$\; \label{algo:basicGen:genpnt}
     \If{$(x,y) \in [0,1]^d$}{ \label{algo:basicGen:check}
       $w=U(0,sp_1)$\;
       $h=U(0,sp_2)$\;
       $\mathcal{G}= \mathcal{G} \cup \{Box(x-w/2,y-h/2,w,h)\}$\;
       $i \leftarrow i + 1$\;}
      } \label{algo:basicGen:loop2}
\Return $\mathcal{G}$
\caption{$G_*()$: basic algorithm for the first five generators.}
\label{algo:basicGen}
\end{algorithm}


\subsection{Uniform}
\begin{figure}[t]
    \centering
    \subfigure[Uniform distribution]{\includegraphics[width=0.3\textwidth]{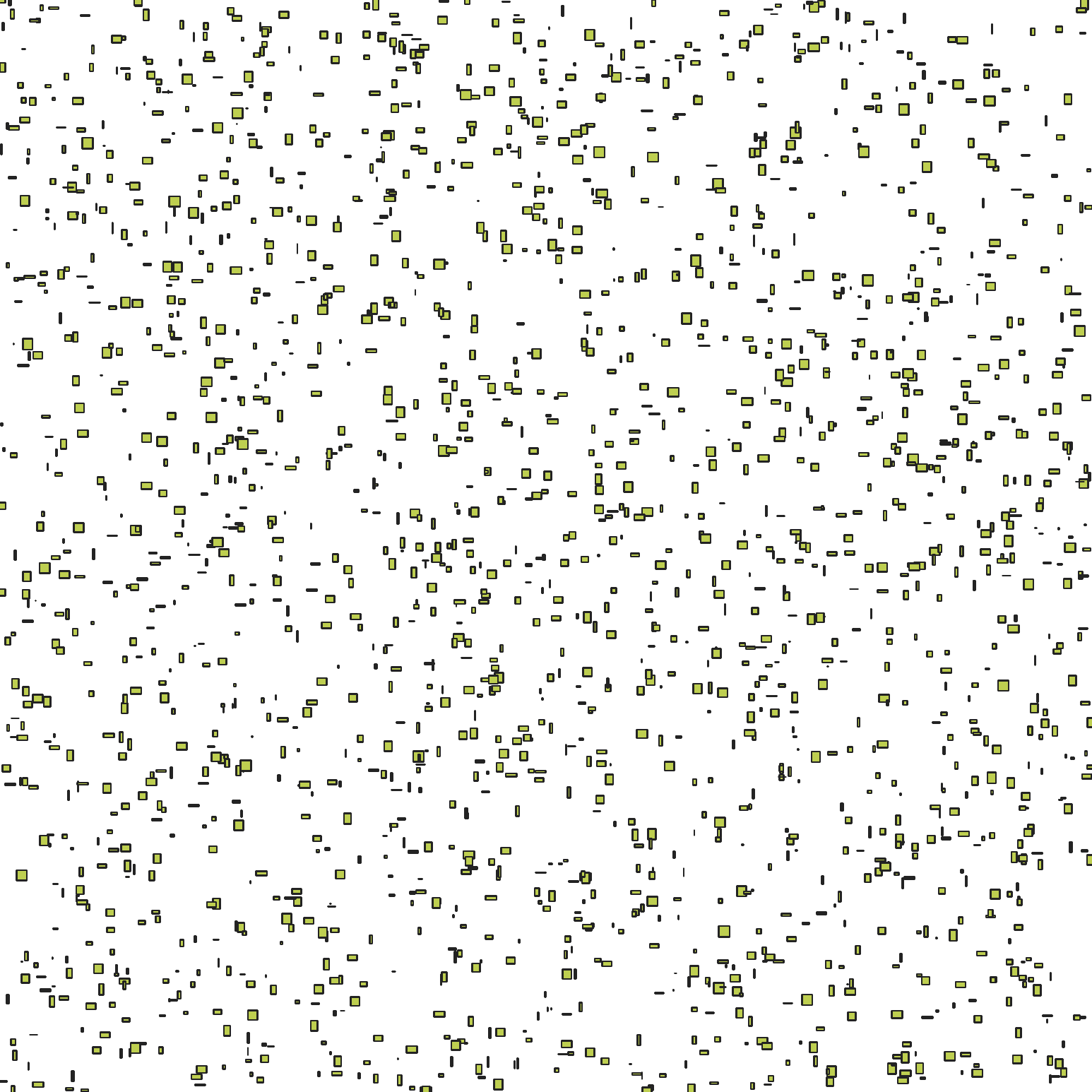} \label{fig:uniform}}
    \subfigure[Diagonal distribution]{\includegraphics[width=0.3\textwidth]{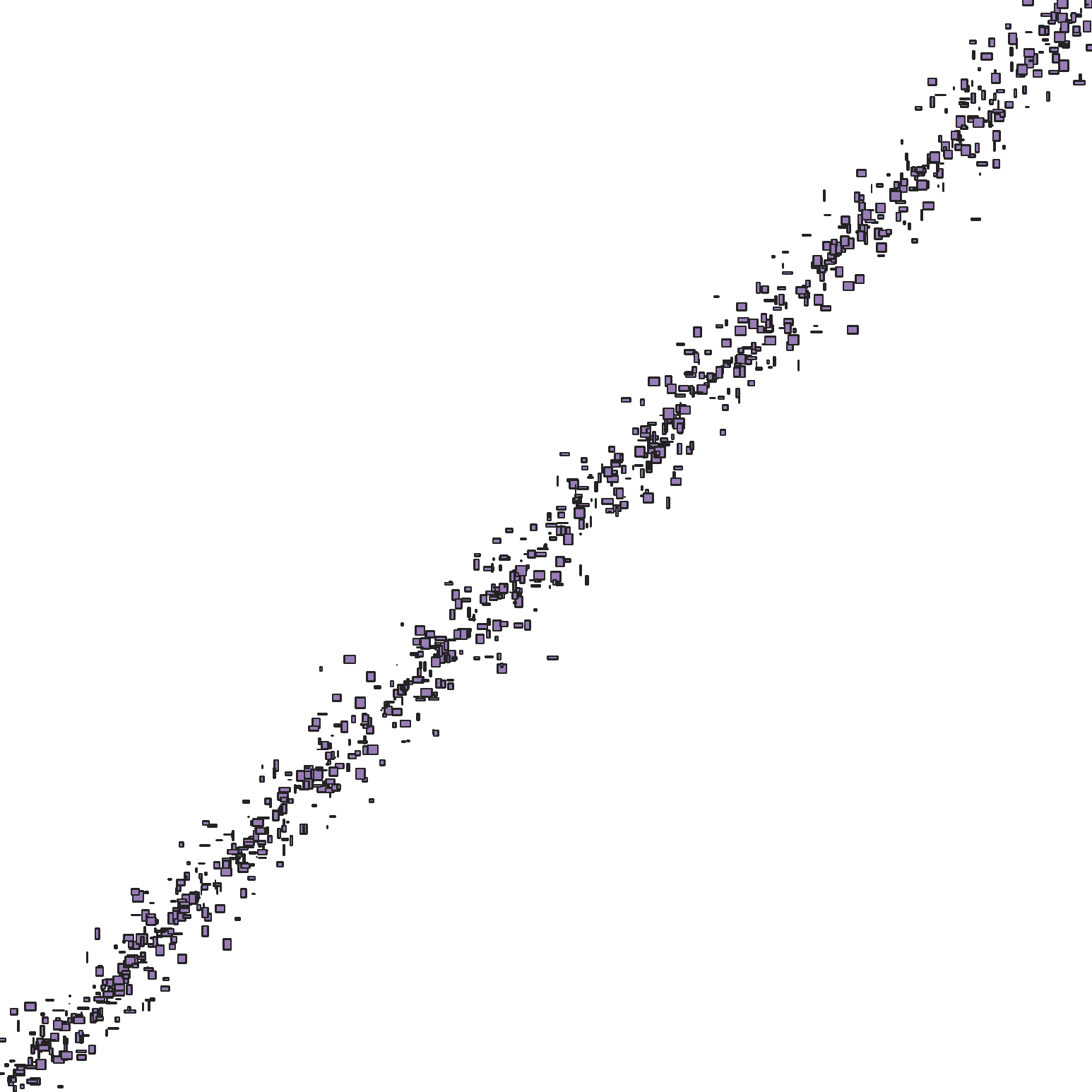} \label{fig:diagonal}}
    \subfigure[Gaussian distribution]{\includegraphics[width=0.3\textwidth]{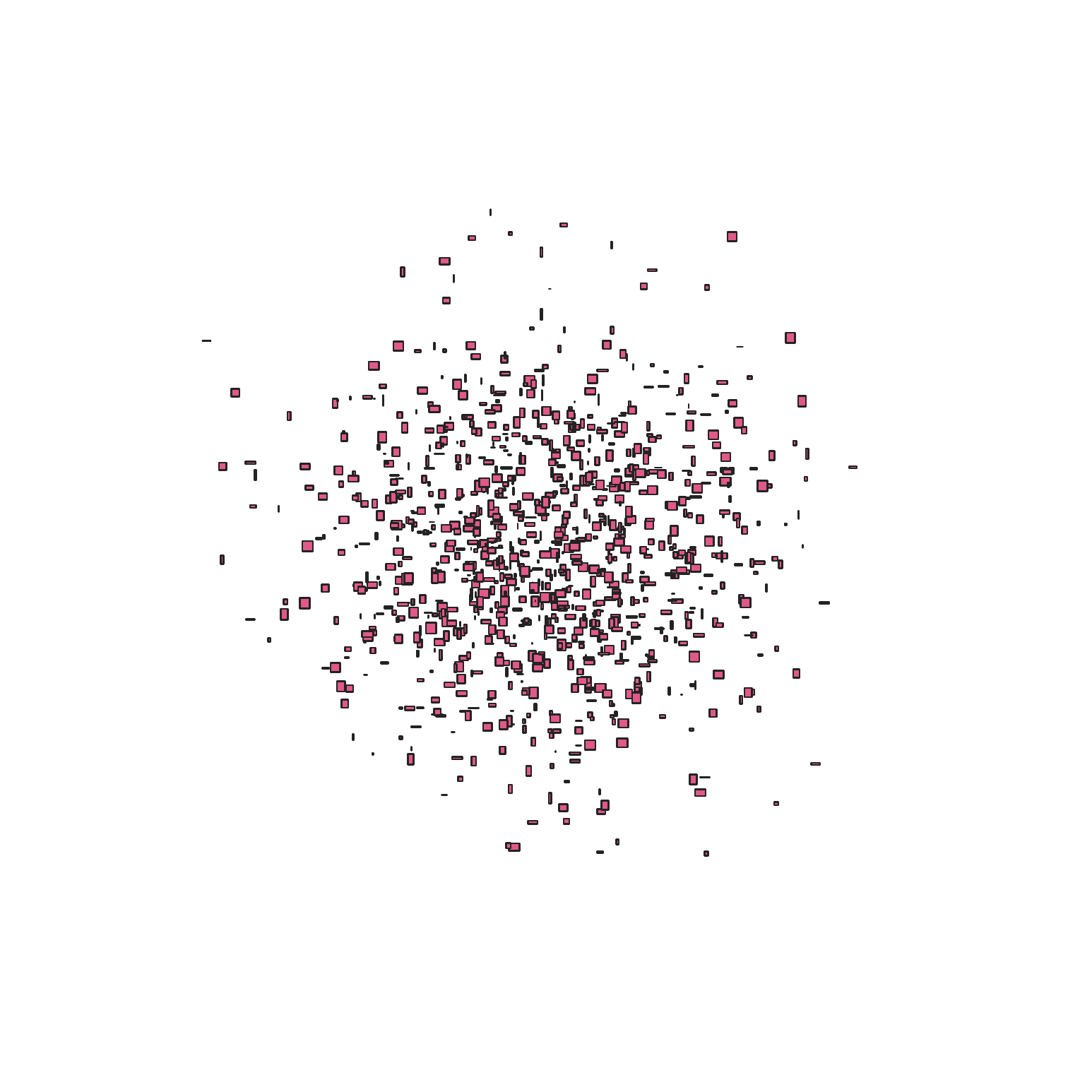} \label{fig:gaussian}}
    \caption{Examples of the first three distributions}
    \label{fig:examples1}
\end{figure}

In the uniform distribution, points are generated randomly inside the reference space $[0,1]^d$ as shown in Figure~\ref{fig:uniform}. This distribution models non-skewed data such as data in suburban areas.
No additional specific parameters are needed for this generator. The point $(x,y)$ is generated using the following equations.

\begin{equation}
    {\sc GenPNT_{uni}()}=\left(x=U(0,1), y=U(0,1)\right)
\end{equation}

\subsection{Diagonal}
\label{ssec:dia}
The diagonal point generator generates points that are concentrated around the diagonal line $x=y$ as illustrated in Figure~\ref{fig:diagonal}. This distribution models real data concentrated around a line such as a river bank or a highway. The affine transformation, described later, can be used to arbitrarily rotate this line.
This generator takes two additional parameters $perc$ and $buf$, where $perc\in[0,1]$ is the percentage (ratio) of the points that are exactly on the line, and $buf\in[0,1]$ is the size of the buffer around the line where additional points are scattered. The additional points are scattered according to a normal distribution. Algorithm~\ref{algo:diagonal} illustrates the generation of a point using the diagonal distribution. Line~\ref{algo:diagonal:if} decides with a probability $perc$ to generate a point exactly on the diagonal. Otherwise, with probability $1-perc$, it generates a point that is shifted with a distance $d$ from the center. The distance is generated from the normal distribution $N(0,buf/5)$ which has almost a $99\%$ probability in generating a number in the range $[-buf,+buf]$. This distance is then divided by $\sqrt{2}$ to calculate the orthogonal offset for $x$ and $y$. Finally, the point location is generated.

\begin{algorithm}[htb]
\SetAlgoLined
\eIf{$Bernoulli(perc) = 1$} { \label{algo:diagonal:if}
  $x=y=U(0,1)$\;
}{
  $c=U(0,1)$\;
  $d=N(0,buf/5)$\;
  $x=c+d/\sqrt{2}$\;
  $y=c-d/\sqrt{2}$\;
}
\Return $(x,y)$\;
\caption{{\sc GenPNT}$_{dia}$ ($perc$, $buf$)}
\label{algo:diagonal}
\end{algorithm}


\subsection{Gaussian}
In the Gaussian distribution, the points are concentrated around the center of the input space $(0.5, 0.5)$ as illustrated in Figure~\ref{fig:gaussian}. This distribution can model real data concentrated around a point such as a metro area.
The coordinates follow a normal distribution $x,y\sim N(0.5,0.1)$. This ensures that almost $99\%$ of the points fall in the box $[0,1]^d$. If points are generated outside that space, the loop in Algorithm~\ref{algo:basicGen} will drop the point and generate another one. This might make the data distribution is a bit different from original Gaussian distribution. In summary, the Gaussian point generator follows the following formula.

\begin{equation}
    \textsc{GenPNT}_{Gaussian}=(x=N(0.5,0.1), y=N(0.5,0.1))
\end{equation}


\subsection{Sierpinski triangle}
\begin{figure}[t]
    \centering
    \subfigure[Sierpinski distribution]{\includegraphics[width=0.3\textwidth]{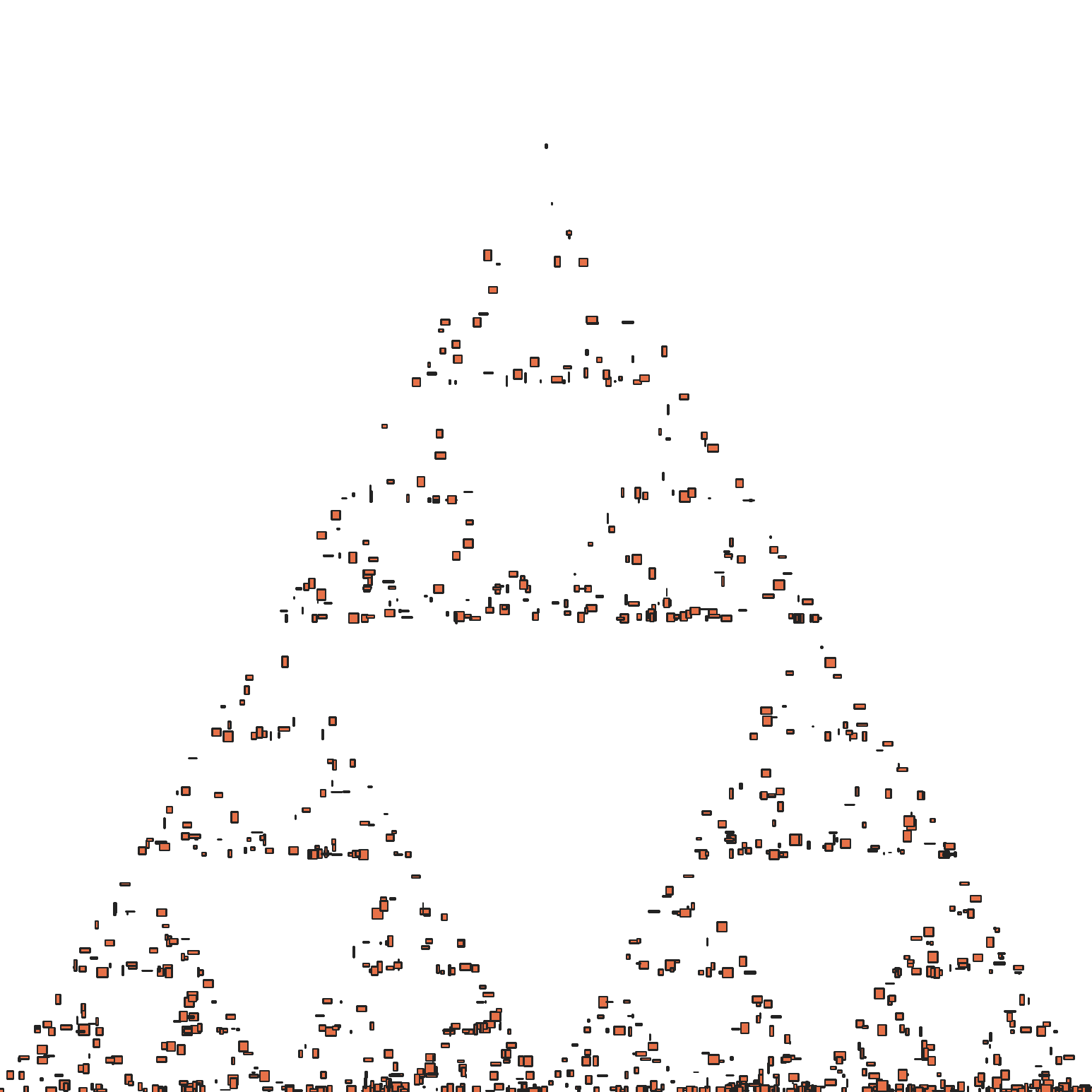} \label{fig:sierpinsky}}
    \subfigure[Bit distribution]{\includegraphics[width=0.3\textwidth]{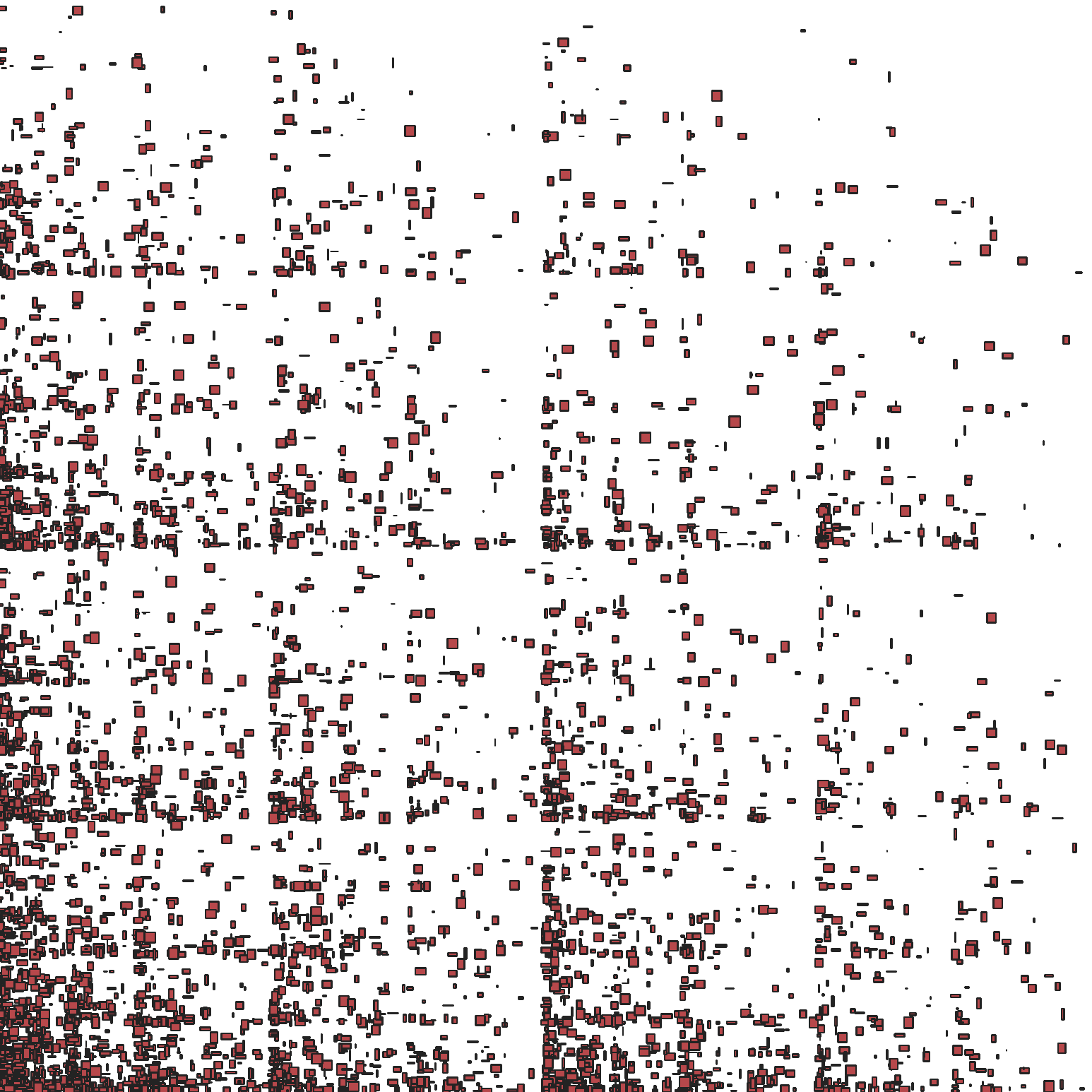} \label{fig:bit}}
    \subfigure[Parcel distribution]{\includegraphics[width=0.3\textwidth]{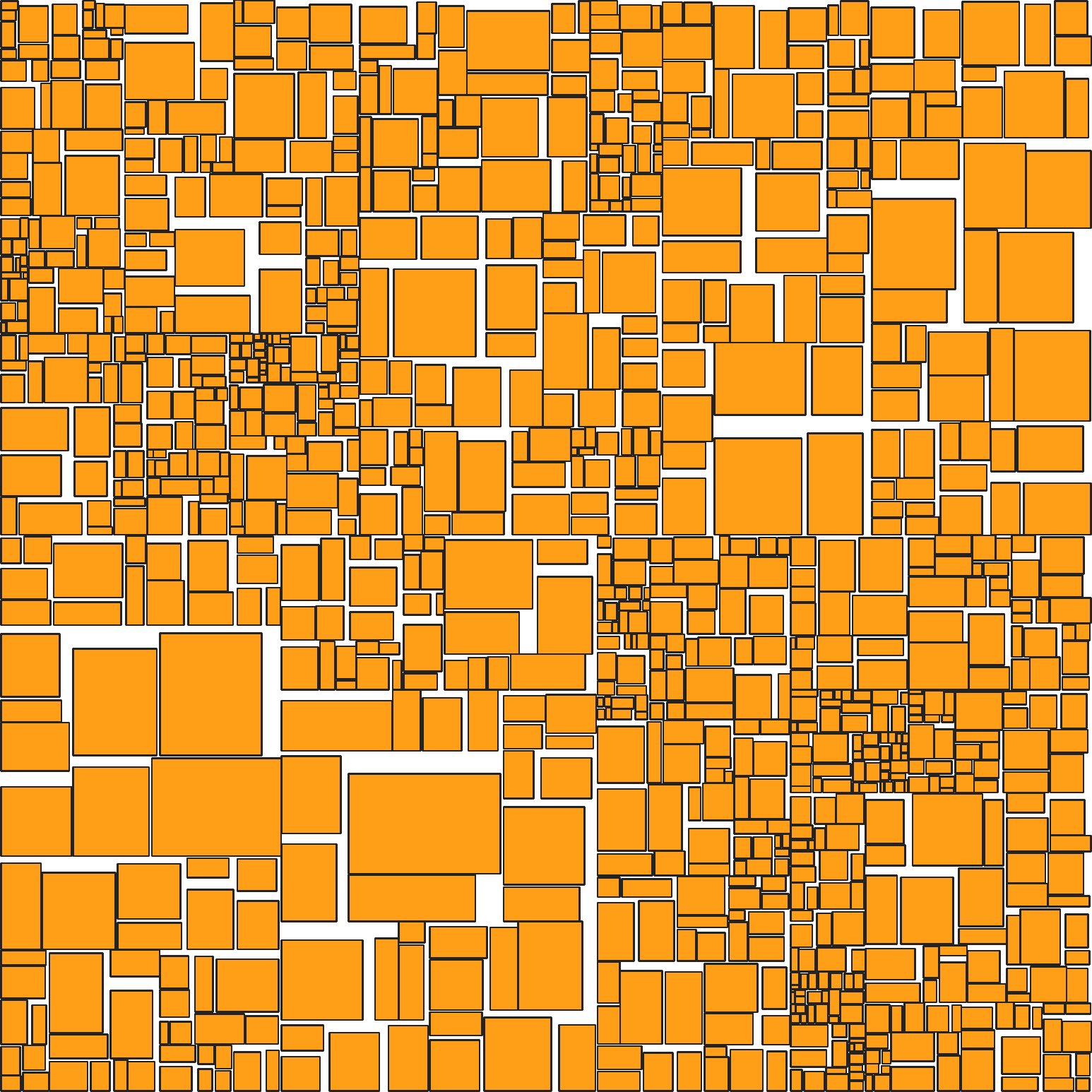} \label{fig:parcel}}
    \caption{Examples of the last three distributions}
    \label{fig:examples2}
\end{figure}

In this case, the skewed distribution is obtained by applying a rule for generating points that belong to a fractal (the Sierpinski's triangle)~\cite{BS88}. Figure~\ref{fig:sierpinsky} gives an example of this data. This pattern could be found in many real model such as cellular automata or motors. This rule is based on an iterative approach such that the generation of the next point of the set depends on the current point and a random function.
The function $\textsc{GenPNT}_{sie}(pnt, i)$ has two specific parameters: the previous point of the iteration $pnt$ and the iteration variable $i$. It generates a two-dimensional point at each iteration as follows:
\begin{align}
& \textsc{GenPNT}_{sie}(pnt, i) =
\left\{
\begin{array}{ll}
(0.0, 0.0) & \textrm{if } i = 0 \\
(1.0, 0.0) & \textrm{if } i = 1 \\
(1/2, \sqrt{3}/2) & \textrm{if } i = 2 \\
\textsc{MiddlePoint}(pnt, (0.0, 0.0)) & \textrm{if } i > 2 \wedge \textsc{Dice}(5) \in \{ 1, 2 \}\\
\textsc{MiddlePoint}(pnt, (1.0, 0.0))  & \textrm{if } i > 2 \wedge \textsc{Dice}(5) \in \{ 3, 4 \}\\
\textsc{MiddlePoint}(pnt, (1/2, \sqrt{3}/2)) & \textrm{if } i > 2 \wedge \textsc{Dice}(5) = 5\\
\end{array}
\right.
\nonumber
\end{align}
where $\textsc{Dice}(5)=\lfloor U(0,5) \rfloor+1$ is a random function producing a number between 1 and 5 and $\textsc{MiddlePoint}(pnt_1, pnt_2)$ computes the middle point between two points, $pnt_1$ and $pnt_2$. 
Notice that, the first three points are the corners of the triangle; the successive points are generated starting from the current point $pnt$ and computing the middle point between $pnt$ and one of the vertices of the triangle chosen according to the random function $\textsc{Dice}(5)$. The vertices of the base are chosen with a probability of $2/5$, the other vertex with a probability of $1/5$. Figure~\ref{fig:sierpinsky} shows an example of the resulting set.

\subsection{Bit distribution}
Another approach for generating a skewed point dataset is to introduce a rule for generating the coordinates of the points by assigning higher probability to a subset of coordinates. For instance in the Bit distribution, the point coordinates are generated as a bit string of a fixed length where each bit is set with a fixed probability $p\in[0,1]$. This generator takes two parameters, $p$ and $digits$, where $p$ represents a fixed probability of setting each bit independently to $1$ and $digits$ represents the number of binary digits after the fraction point.
It generates a point in a higher dimensional space by setting each dimension independently in the same method as shown below:
\begin{equation}
\textsc{GenPNT}_{bit}(p, digits) = 
(\textsc{Bit}(p, digits) , \textsc{Bit}(p, digits) )
\nonumber
\end{equation}
where $\textsc{Bit}(p, digits)$ generates a real number between $0.0$ and $1.0$ as shown in the following Algorithm~\ref{algo:bit},
\begin{algorithm}[htb]
\SetAlgoLined
\KwIn{$p, digits$}
\KwResult{Real number: $n$}
$n \leftarrow 0.0$; $i \leftarrow 0$\;
\For{$i = 1$ to $digits$}
    {
    $c\leftarrow$Bernoulli(p)\;
    $n \leftarrow n + c /2^i$\;
    }
\Return $n$
\caption{$\textsc{Bit}(p, digits)$}
\label{algo:bit}
\end{algorithm}

\subsection{Parcel distribution}
This generator directly generates rectangles according to the parcel distribution. The parcel distribution generates geometries that represent boxes of different sizes as illustrated in Figure~\ref{fig:parcel}. This distribution can model land sections delineated in urban areas. 

In addition to the cardinality $card$ of the generated dataset, the parcel generator takes two specific parameters $r$ and $d$, where:
\begin{itemize}
    \item $r\in[0,0.5]$ is the minimum tiling range for splitting a box. $r=0$ indicates that all the ranges are allowed while $r=0.5$ indicates that a box is always split into half.
    \item $d\in[0,1]$ is the dithering parameter that adds some random noise to the generated rectangles. $d=0$ indicates no dithering and $d=1.0$ indicates maximum dithering that can shrink rectangles down to a single point.
\end{itemize}

Algorithm~\ref{algo:parcel} describes how the parcel generator works. The first loop in lines~\ref{algo:parcel:loop1start}-\ref{algo:parcel:loop1end} repetitively splits the reference space $[0,1]^d$ along one of the two axes $x$ and $y$. It always splits along the longer axis of the given box. This loop runs $card-1$ times as it generates one new box at each iteration.

The second loop in lines~\ref{algo:parcel:loop2start}-\ref{algo:parcel:loop2end} adds the dithering effect by shrinking each box with a ratio $1-U(0,d)$. This means that if $d=0$, the ratio will always be equal to $1.0$ which means no shrinking. If $d=1.0$, the ratio can reach up-to 1.0 which shrinks the boxes all the way to a single point.

\begin{algorithm}[htb]
\SetAlgoLined
\KwIn{$card, d=2, r, d$}
\KwResult{Set of geometries: $\mathcal{G}=\{ geom \}$}
Initialize the random number generator\; \label{algo:parcel:init1}
$\mathcal{G} \leftarrow \{Box(0, 0, 1.0, 1.0)\}$\; \label{algo:parcel:init2}
\While{$|\mathcal{G}| < card$} {\label{algo:parcel:loop1start}
  $b\leftarrow$ $\mathcal{G}$.dequeue\;
  \eIf{b.width > b.height}{
    splitSize = b.width $* U(r,1-r)$ \;
    $b_1$=Box(b.x, b.y, splitSize, b.height)\;
    $b_2$=Box(b.x + splitSize, b.y, b.width - splitSize, b.height)\;
  }{
    splitSize = b.height $* U(r,1-r)$ \;
    $b_1$=Box(b.x, b.y, b.width, splitSize)\;
    $b_2$=Box(b.x, b.y + splitSize, b.width, b.height - splitSize)\;
  }
  $\mathcal{G}$.enqueue($b_1$);
  $\mathcal{G}$.enqueue($b_2$);
} \label{algo:parcel:loop1end}
\For{$b \in \mathcal{G}$} { \label{algo:parcel:loop2start}
  $b.width =b.width \cdot (1-U(0,d))$\;
  $b.height = b.height \cdot (1-U(0,d))$\;
} \label{algo:parcel:loop2end}
\Return $\mathcal{G}$
\caption{$G_{parcel}(r,d)$: Generated boxes of the parcel distribution}
\label{algo:parcel}
\end{algorithm}


\section{Post Transformations}
This section describes two methods that can giver further flexibility on customizing the generated data, transformation and compounding. The transformation method applies a simple affine transformation on the generated geometries. The compounding method combines several datasets by simply unifying all their geometries.

\subsection{Affine Transformation}
\label{sec:affine}
The generators described above are designed to be very simple on purpose. Instead of complicating each generator, most of the customization is moved to this step. This step simply applies a standard affine transformation to all the generated geometries. In the case of points, the transformation is applied to the coordinates of the point. For rectangles, the transformation is applied to the two opposite corners, i.e., the lower left and upper right corners. This ensures that the rectangle remains orthogonal even after rotation which is usually desired in data structures and algorithms that deal with bounding boxes.

An affine transformation is defined by a fixed-size matrix. For two-dimensional data, the affine transformation transforms a point $(x,y)$ to a transformed point $(x',y')$ according to the following equation.

\begin{equation}
\left[ \begin{array}{c}
    x'\\
    y'\\
    1
\end{array} \right] =
\left[ \begin{array}{ccc}
    a_1 & a_2 & a_3  \\
    a_4 & a_5 & a_6 \\
    0 & 0 & 1
\end{array}
\right] \left[
\begin{array}{c}
     x \\
     y \\
     1
\end{array}
\right]
\end{equation}

where $a_1\cdots a_6$ are the parameters of the affine transformation. This formula could be modified to work with higher dimensional data.

\subsection{Compound Datasets}
The final stage is to combine several datasets of either the same distribution but different parameters, different distributions, or with different transformation matrices.
Show some examples of how combining datasets can generate new interesting datasets.

\begin{figure}[ht]
\centering
\includegraphics[width=0.6\textwidth]{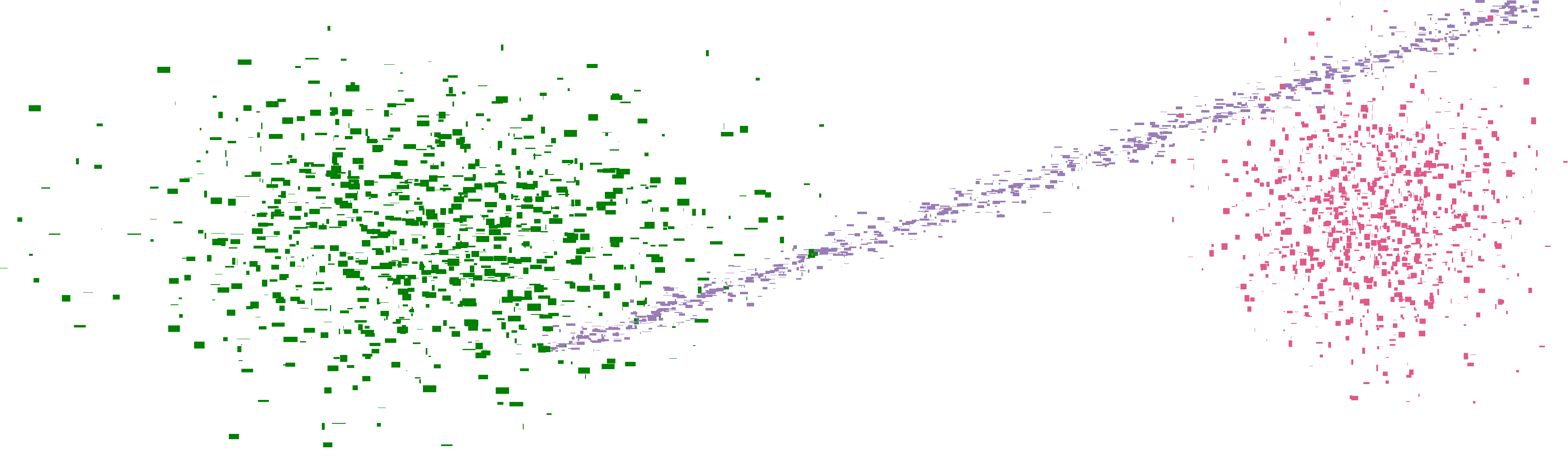}
\caption{Example of compound dataset obtained by combining two different Gaussian distributions and one diagonal distribution.}
\label{figure.compound_distribution}
\end{figure}


\begin{table}[t]
    \centering
    \begin{tabular}{l|r r r r r r | r r r r r r}
        \hline
         Distribution ID & $card$ & $d$ & $sp_1$ & $sp_2$ & $sp_3$ & $sp_4$ & $a_1$ & $a_2$ & $a_3$ & $a_4$ & $a_5$ & $a_6$ \\
         \hline
         Uniform (Figure~\ref{fig:uniform}) & 1000 & 2 & 0.02 & 0.02 & & & 1 & 0 & 0 & 0 & 1 & 0 \\
         Diagonal (Figure~\ref{fig:diagonal})& 1000 & 2 & 0.01 & 0.01 & 0.2 & 0.1 & 1 & 0 & 0 & 0 & 1 & 0 \\
         Gaussian (Figure~\ref{fig:gaussian})& 2000 & 2 & 0.1 & 0.1 & & & 1 & 0 & 0 & 0 & 1 & 0\\
         Sierpinski (Figure~\ref{fig:sierpinsky})& 1000 & 2 & 0.01 & 0.01 & & & 1 & 0 & 0 & 0 & 1 & 0\\
         Bit (Figure~\ref{fig:bit})& 5000 & 2 & 0.01 & 0.01 & 0.3 & 10 & 1 & 0 & 0 & 0 & 1 & 0\\
         Parcel (Figure~\ref{fig:parcel})& 1000 & 2 & 0.2 & 0.2 & & & 1 & 0 & 0 & 0 & 1 & 0\\
         \hline
    \end{tabular}
    \caption{Identifiers for the six sample datasets shown in this paper. For simplicity, all of them use the identity transformation (affine) matrix}
    \label{tab:datasets}
\end{table}

\subsection{Identifying Datasets}
Based on the proposed method, each generated simple datasets, i.e., not compound, can be identified using a fixed vector. This vector is the one illustrated in Figure~\ref{fig:overview} and it contains the generator model $G_*$, the common parameters $card$ and $d$, the specific parameters $sp_1\cdots sp_n$, and the affine transformation matrix parameters $a_1\cdots a_6$.
Researchers who use the generators described in this paper can simply list all these parameters in a table to allow other researchers to generate datasets of the same characteristics. For example, Table~\ref{tab:datasets} identifies the six sample datasets illustrated in Figure~\ref{fig:examples1} and~\ref{fig:examples2}.

\begin{figure}[t]
    \centering
    \includegraphics[width=\linewidth]{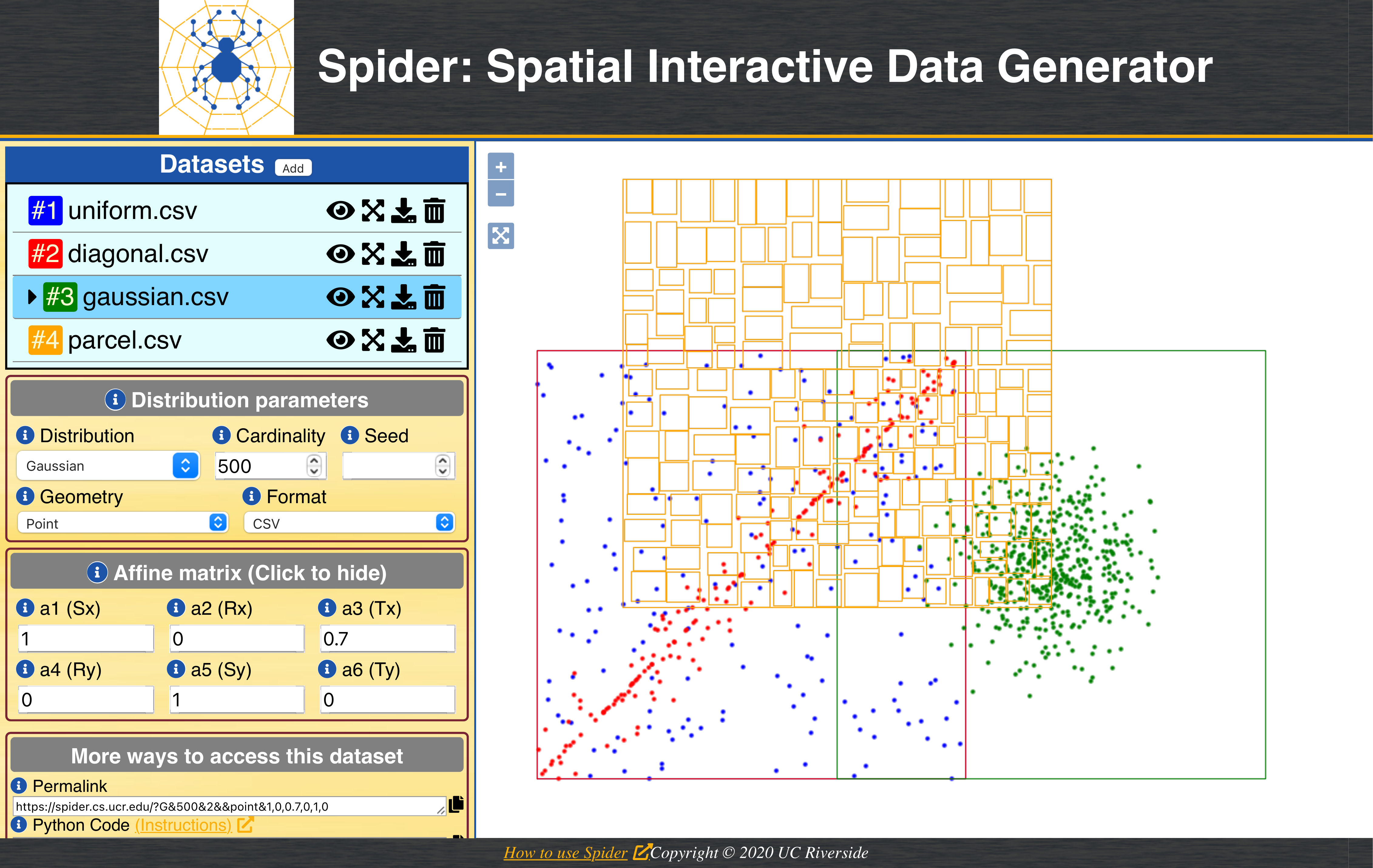}
    \caption{Spider Web: Online service for spatial data generation [\url{https://spider.cs.ucr.edu}]}
    \label{fig:spider-screenshot}
\end{figure}

\section{Spider: Web-based Spatial Data Generator}
The proposed spatial data generator is available as open source~\cite{github-sdg} in Python for users with basic programming skills. In order to serve a wider range of users, Spider Web~\cite{katiyar2020spiderweb, webapp-sdg} was proposed as a web-based spatial data generator based on the proposed design. Figure~\ref{fig:spider-screenshot} shows a simple example of data generation using Spider Web. Users can choose the distribution type, parameters, and affine matrix in the panel on the left. The visualization immediately reflects how the dataset will look like. Users can then download the selected dataset with an arbitrary size in standard formats, CSV, WKT, or GeoJSON. Furthermore, as illustrated in Figure~\ref{fig:spider-screenshot}, users can generate multiple datasets and visually compare them. Finally, Spider Web provides a permalink for any generated dataset that can be shared between team members or researchers to promote the reproducibility of results. With these comprehensive features, Spider Web aims to be a standard tool to improve the reproducibility of experiments in the spatial research community.

\section*{Acknowledgment}
This work is supported in part by the National Science Foundation (NSF) under grants IIS-1838222 and CNS-1924694 and
by the USDA National Institute of Food and Agriculture, AFRI award number A1521.

\bibliographystyle{unsrt}  
\bibliography{references}

\end{document}